\documentclass{article}
\usepackage{graphicx} 

\title{Fast TTC Computation}
\author{Irene Aldridge}
\date{March 17, 2024}

\begin{document}
\maketitle
\begin{abstract}
    This paper proposes a fast Markov Matrix-based methodology for computing Top Trading Cycles (TTC) that delivers O(1) computational speed, that is speed independent of the number of agents and objects in the system. The proposed methodology is well suited for complex large-dimensional problems like housing choice. The methodology retains all the properties of TTC, namely, Pareto-efficiency, individual rationality and strategy-proofness.  
\end{abstract}
\section{Introduction}
\textbf{}The Top Trading Cycles Algorithm due Gale and \cite{SHAPLEY197423} is a Pareto-efficient way to allocate objects based on agents preferences. Working with strict preferences of agents over objects, the algorithm estimates the allocations that are individually-rational, envy-free and strategy-proof. Perhaps the only drawback of the TTC algorithm is that it is computationally slow. \cite{SabanSethuraman2013} showed that a conventional TTC algorithm for a system with strict preferences runs in O(n log n) time, where n is the number of agents and objects in the system. 

This paper proposes an optimization of the TTC algorithm. Specifically, to reduce the computational burden of TTC, we propose to map TTC onto a Markov Matrix, which can be solved efficiently in O(1), i.e., independently of the number of agents or objects in the system (\cite{SunEtAl2020}). 

Converting preferences to a Markov Matrix allows us to quickly identify and eliminate cycles, and assign proper order to the agents to choose the best objects available to them at the time of their choice. 

\section{Methodology}

\subsection{Methodology Overview}

A Markov Matrix, also known as a stochastic matrix, is a $n$ x $n$ matrix summarizing probabilities of a given agent moving from state $i$ to state $j$. This probability is recorded in the cell $A_{ij}$. All the columns are normalized to sum up to 1 to reflect the closed state of the system: all paths leaving a given state are reflected in the matrix. 

The solution to a Markov Matrix produces stable-state inferences: the steady-state probabilities of finding an agent in a given state. These inferences are free of cycles, since cycles or periodic movements are not stable and are not reflected in the steady-state inferences. A Perron-Frobenius solution to finding the steady-state inferences then delivers the $O(1)$ estimate via singular value calculations (\cite{SunEtAl2020}). 

In this research, we propose harnessing the power of Markov Matrices by expressing agent preferences in a Markov-like model. We start with agents $i = \{1,..., I\}$ and assign as their “destinations” objects $j = \{1,..., N\}$. For each transfer from state $i$ to state $j$, we assign a "probability" $P_{ij}$ that reflects agent $i$’s preferences over objects $\{j\}$. Specifically, we assign probability $1$ to transfer $ij$ if $j$ is the number 1 choice for agent $i$. Next, given the rank choice $R_{ij}$ of agent $i$ over the object $j$, we assign probability $P_{ij} = (N-R_{ij})/N$ to the transition from $i$ to $j$. 

Once our Markov Matrix is complete, we next estimate the steady-state probabilities as the first singular vector of the matrix, following Perron-Frobenius Theorem (see, for example, \cite{BigDataScienceInFinance}). The relative ordering of the coefficients in the first singular vector follows cyclicality of the agents and their preferences: the lower the coefficient, the less stable is the transition, the more likely it is to be in the cycle, the more priority it is given in our algorithm described below.

\subsection{A Toy Example 1}
\textbf{}
To illustrate the idea, consider the following matrix of preferences over four objects {1, 2, 3, 4}. 

P =  	[[1 2 3 4]
 	[4 1 3 2]
 	[2 1 4 3]
 	[1 4 3 2]]

The TTC graph corresponding to P looks as follows: 
\begin{enumerate}
    \item Among the highest-level preferences, identify cycles by connecting agents with their best-choice objects (Figure 1). 
    \item Eliminate objects that are either self-referencing or in a cycle (Figures 2, 3 and 4).
    \item Repeat the process for the remaining agents and objects (Figure 5).
\end{enumerate}

\begin{figure}
    \centering
    \includegraphics[width=0.5\linewidth]{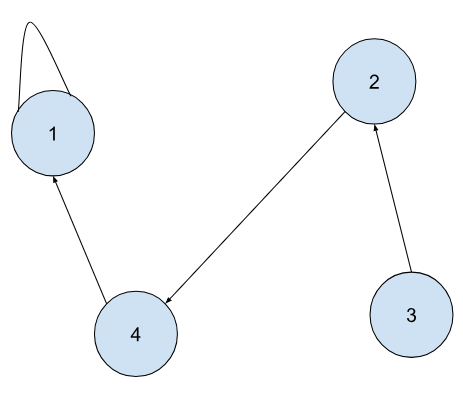}
    \caption{Step 1: identify the highest-level preferences. }
    \label{fig:enter-label}
\end{figure}

\begin{figure}
    \centering
    \includegraphics[width=0.5\linewidth]{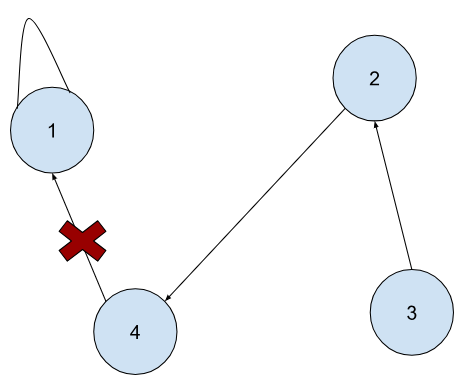}
    \caption{Step 2: observe that Agent 1 prefers object 1, eliminate Agent 1 and object 1.}
    \label{fig:enter-label}
\end{figure}

\begin{figure}
    \centering
    \includegraphics[width=0.5\linewidth]{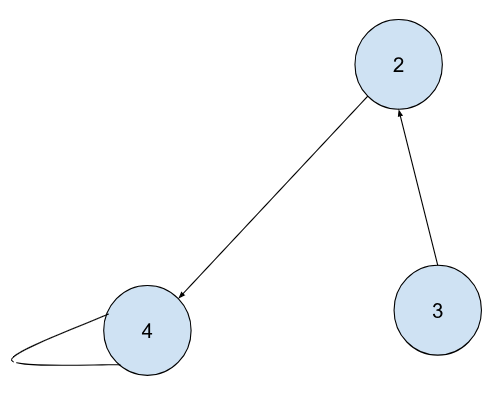}
    \caption{Step 3. Observe that, after object 1 exists, Agent 4 prefers object 4. }
    \label{fig:enter-label}
\end{figure}

\begin{figure}
    \centering
    \includegraphics[width=0.5\linewidth]{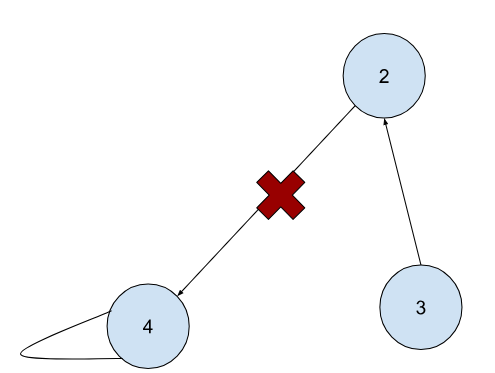}
    \caption{Step 4. Agent 4 and object 4 exit the system. }
    \label{fig:enter-label}
\end{figure}

\begin{figure}
    \centering
    \includegraphics[width=0.5\linewidth]{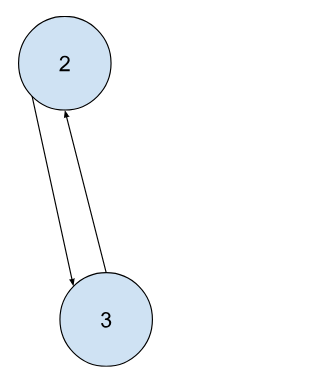}
    \caption{Step 5. Agent 2 updates his preferences because objects 1 and 4 are no longer available. Agent 2 receives object 3 and Agent 3 receives object 2. }
    \label{fig:enter-label}
\end{figure}

The resulting "classical" TTC Solution is:

Agent 1 → 1
Agent 2 → 3
Agent 3 → 2
Agent 4 → 4

In this paper, we propose to:
\begin{enumerate}
    \item Summarize preferences using a novel weighted directed graph approach.
    \item Use Perron-Frobenius Theorem to efficiently compute the cycles in $O(1)$ time.
\end{enumerate}

Thus, a graph for the preference matrix $P$ can be represented as graph $G$, with the first choice as 1, the second choice as $(N-1)/N$, … and the last choice as $1/N$:

$P =$
        1: $1\succ 2\succ 3\succ 4$ \\
 	2: $4\succ 1\succ 3\succ 2$ \\
 	3: $2\succ 1\succ 4\succ 3$ \\
 	4: $1\succ 4\succ 3\succ 2$
    
G = 	[[1, ¾, ½ , ¼],
	[¾, ¼, ½, 1].
[¾, 1, ¼, ½],
[1, ¼, ½, ¾]]

The matrix G documents agent preferences over objects in a linear fashion. It can be thought of as the choice each agent i would make given the opportunity to pick any of the 4 states, which in TTC are commonly referred to as objects. Thus, Agent 1 would much prefer to stay with state 1, then state 2, state 3, and state 4, in that order. At the same time, Agent 1 would much rather move to state 4, then state 1, state 3 and state 2. In the context of this discussion, we use a linear probability scale, and the resulting outcome is consistent with TTC. We can, however, generalize the approach to another probability distribution. 

After summarizing each agent’s preferences over objects on a 0-1 scale, we normalize matrix G to ensure all columns add up to 1. To do so, we sum up the columns and divide each element by their respective sum of the column:

$\sum_i G_{ij}=$  [3.5    2.25    1.75     2.5 ]

normalized matrix =  	[[0.28571429 0.33333333 0.28571429 0.1]

    [0.21428571 0.11111111 0.28571429 0.4       ]
    
    [0.21428571 0.44444444 0.14285714 0.2       ]
    
    [0.28571429 0.11111111 0.28571429 0.3       ]]

Next, we apply the singular value decomposition and consider the first singular vector:
$V[0]= [-0.67952481 -0.43424211 -0.33978586 -0.48396838]$

The smallest values in this vector represent the cycles and are removed in their respective order:
Thus, the first agent is the first to obtain his wish (object 1), 
Agent 4 is next, taking out object 4, 
Agent 2 is next with object 3, followed by 
Agent 3 with object 2.

The process delivers results identical to the conventional TTC approach shown above.

\subsection{A Toy Example 2}
\textbf{}

P =  [[1 2 3 4 5]
 [5 4 1 3 2]
 [2 1 5 4 3]
 [1 5 4 3 2]
 [2 3 5 4 1]]

The TTC algorithm works as follows:

Figure . Example 2, TTC Round 1

Cycles 1 and 2 $\leftrightarrow$ 5 are eliminated, leaving agents 3 and 4.

In the next round, Agent 4 prefers 4, and exits, leaving Agent 3 with object 3.

We can replicate the same process with our methodology. Create a weighted graph G as follows: 

$G = 	[[1, 4/5, 3/5, 2/5, 1/5],$

	$[3/5, 1/5, 2/5, 4/5, 1 ],$
 
	$[4/5, 1, 1/5, 2/5, 3/5 ],$
 
	$[1, 1/5, 2/5, 3/5, 4/5 ],$
 
	$[1/5, 1, 4/5, 2/5, 3/5 ]]$

In other words, 
$G = 	[[1, 0.8, 0.6, 0.4, 0.2],$

$[0.6, 0.2, 0.4, 0.8, 1], $

$[0.8, 1, 0.2, 0.4, 0.6], $

$[1, 0.2, 0.4, 0.6, 0.8], $

$[0.2, 1, 0.8, 0.4, 0.6]]$

$\sum_i G_{ij}= $  [3.6  3.2  2.4  2.6  3.2]

Normalized matrix =  [[0.27777778 0.25       0.25       0.15384615 0.0625    ]

 [0.16666667 0.0625     0.16666667 0.30769231 0.3125    ]

 [0.22222222 0.3125     0.08333333 0.15384615 0.1875    ]

 [0.27777778 0.0625     0.16666667 0.23076923 0.25      ]

 [0.05555556 0.3125     0.33333333 0.15384615 0.1875    ]]

$V[0] = [-0.53588536 -0.47158305 -0.35029467 -0.38264279 -0.47044069]$

Here, the first agent makes his choice, followed by the second, the fifth, the fourth and the third agents:
Agent 1 → 1
Agent 2 → 5
Agent 5 → 2
Agent 4 → 4
Agent 3 → 3

Our methodology delivers the same assignment as TTC, but is very fast and infinitely scalable to very large systems. Since we are using TTC as the core methodology, our algorithm delivers Pareto-efficient, strategy-proof and individually-rational allocations, just like TTC does. 

\subsection{Large-Sample Extension}
One of the advantages of SVD methodology is that it can be readily extended to matrices with an arbitrarily large number of data observations. Unlike its cousin Principal Component Analysis (PCA), SVD does not impose any restrictions on the shape of matrices. 

In a particular version of the algorithm proposed in this article, the number of agents needs to equal the number of objects, making this method suitable for problems like Housing Assignment (\cite{ABDULKADIROGLU1998}, \cite{AbdulkadirogluSonmez2003}) and others where the number of agents can be very large and the number of distinct available objects can be equal. Furthermore, unlike recently-proposed algorithms like \cite{BogomolnaiaMoulin2004} and \cite{BOGOMOLNAIA2001295}, SVD does not admit randomization. To many agents, randomization may seem unfair and thus inducing envy. The proposed methodology is, therefore, more optimal.

\section{Conclusion}
\textbf{}
The proposed computation of TTC algorithm delivers a fast solution to an object allocation problem, regardless of the size of the data at hand. It can be efficiently used to allocate extremely large arrays of agents, each with well-defined multiple preferences, such as public schools to prospective students and so on. The algorithm retains all the properties of TTC, namely Pareto-optimality, strategy-proofness and individual rationality. 

\bibliographystyle{plain}
\bibliography{References}
\end{document}